\documentclass[english,onecolumn,11p]{IEEEtran}

\usepackage[latin9]{inputenc}
\usepackage{geometry}
\usepackage{amsthm}
\usepackage{amsmath}
\usepackage{amssymb}
\usepackage{fixltx2e}
\usepackage{graphicx}
\usepackage{setspace}
\usepackage{esint}
\usepackage{xargs}[2008/03/08]
\makeatletter
  \theoremstyle{plain}
  \newtheorem{thm}{\protect\theoremname}
  \theoremstyle{plain}
  \newtheorem{lem}{\protect\lemmaname}

\usepackage{amsmath}
\usepackage{amssymb}
\usepackage{amsthm}
\usepackage{amsfonts}
\usepackage{epsfig}
\usepackage{latexsym}
\usepackage{url}
\usepackage[hypertexnames=false]{hyperref}
\usepackage{ifthen}
\usepackage[noadjust]{cite}

\makeatother

\usepackage{babel}
\providecommand{\lemmaname}{Lemma}
\providecommand{\theoremname}{Theorem}

\begin{document}

\title{Lower Bounds and Approximations for the Information Rate of the ISI
Channel}

\author{Yair Carmon and Shlomo Shamai\IEEEauthorrefmark{1} %
\thanks{\IEEEauthorrefmark{1} Technion, Israel Institute of Technology. Emails:
yairc@tx.technion.ac.il, sshlomo@ee.technion.ac.il%
}}
\maketitle
\begin{abstract}
We consider the discrete-time intersymbol interference (ISI) channel
model, with additive Gaussian noise and fixed i.i.d. inputs. In this
setting, we investigate the expression put forth by Shamai and Laroia
as a conjectured lower bound for the input-output mutual information
after application of a MMSE-DFE receiver. A low-SNR expansion is used
to prove that the conjectured bound does not hold under general conditions,
and to characterize inputs for which it is particularly ill-suited.
One such input is used to construct a counterexample, indicating that
the Shamai-Laroia expression does not always bound even the achievable
rate of the channel, thus excluding a natural relaxation of the original
conjectured bound. However, this relaxed bound is then shown to hold
for any finite entropy input and ISI channel, when the SNR is sufficiently
high. Finally, new simple bounds for the achievable rate are proven,
and compared to other known bounds. Information-Estimation relations
and estimation-theoretic bounds play a key role in establishing our
results.
\end{abstract}
\global\long\def\Immse{{I_{\mathrm{MMSE}}}}
\global\long\def\Isow{{I_{\textrm{SOW}}}}
\global\long\def\Isl{{I_{\textrm{SL}}}}
\global\long\def\Iach{{\mathcal{I}}}
\newcommandx\SNR[1][usedefault, addprefix=\global, 1=x]{{\mathsf{SNR}_{\textrm{#1}}}}
\global\long\def\Kset{{\mathcal{K}}}
\global\long\def\Mset{{\mathcal{M}}}
\global\long\def\Pset{{\mathcal{P}}}

\section{\label{sec:intro}Introduction and preliminaries}

The discrete-time inter-symbol interference (ISI) communication channel
model is given by,
\begin{equation}
y_{k}=\sum_{i=0}^{L-1}h_{i}x_{k-i}+n_{k}\label{eq:ISI_model}
\end{equation}
where $x_{-\infty}^{\infty}$%
\footnote{We use the standard notation $a_{N_{1}}^{N_{2}}$ for the sequence
$[a_{N_{1}},a_{N_{1}+1},...,a_{N_{2}}]$ with the natural interpretation
when $N_{1}=-\infty$ and/or $N_{2}=\infty$.%
}, is an independent identically distributed (i.i.d.) channel input
sequence with average power $P_{x}=E{x_{0}^{2}}$ and $y_{-\infty}^{\infty}$
is the channel output sequence. The noise sequence $n_{-\infty}^{\infty}$
is assumed to be an i.i.d. zero-mean Gaussian sequence independent
of the inputs, with average power $N_{0}=E{n_{0}^{2}}$, and $h_{0}^{L-1}$
are the ISI channel coefficients. We let $H\left(\theta\right)=\sum_{k=0}^{L-1}h_{k}e^{-jk\theta}$
denote the channel transfer function. For simplicity we assume that
the input, ISI coefficients and noise are real, but all the results
reported in this paper extend straightforwardly to a complex setting.

ISI is common in a wide variety of digital communication applications,
and thus holds much interest from both practical and theoretical perspectives.
In particular, evaluation of the maximum achievable rate of reliable
communication sheds light on the fundamental loss caused by ISI, and
aids in the design of coded communication systems. Since this model
is ergodic, the rate of reliable communication is given by \cite{gray2010entropy},
\begin{equation}
\Iach=\lim_{N\rightarrow\infty}\frac{1}{2N+1}I\left(x_{-N}^{N}\,;\, y_{-N}^{N}\right)\label{eq:Achievable rate}
\end{equation}

When the input distribution is Gaussian, a closed form expression
for $\Iach$ is readily derived by transforming the problem into parallel
channels (cf. \cite{cover2012elements}), and is given by
\begin{equation}
\Iach_{g}=\frac{1}{2\pi}\int_{-\pi}^{\pi}\log\left(1+\frac{P_{x}}{N_{0}}|H(\theta)|^{2}\right)d\theta\label{eq:iid capacity}
\end{equation}
This rate is also the maximum information rate attainable by any i.i.d.
input process --- i.e. the i.i.d. channel capacity. However, in practical
communication systems the channel inputs must take values from a finite
alphabet, commonly referred to as a signal constellation. In this
case no closed form expression for $\Iach$ is known. In lieu of such
expression, $\Iach$ can be approximated or bounded numerically, mainly
by using simulation-based techniques \cite{pfister2001achievable,radosevic2011bounds,rusek2009lower,sadeghi2009optimization,arnold2006simulation,arnold2001information}.

Simple closed form bounds on $\Iach$ present an alternative to numerical
approximation. It is straightforward to show that (see \cite{shamai1996intersymbol}),
\begin{equation}
\Iach\geq I\left(x_{0}\,;\,\sum_{k}{a_{k}y_{-k}}\,|\, x_{-\infty}^{-1}\right)\label{eq:Filter bound}
\end{equation}
where $a_{-\infty}^{\infty}$ is an arbitrary set of coefficients.
Substituting for $y_{k}$ according to the channel model (\ref{eq:ISI_model}),
this bound can be simplified to
\begin{equation}
\Iach\geq I\left(x_{0}\,;\, x_{0}+\sum_{k\geq1}{\alpha_{k}x_{k}}+m\right)\label{eq:Filter bound simplified}
\end{equation}
with the coefficients $\alpha_{k}={\sum_{l}{a_{l}h_{-l-k}}}/{c}$,
$c=\sum_{l}{a_{l}h_{-l}}$ and $m$ a Gaussian RV, independent of
$x_{0}^{\infty}$ with zero mean and variance $Em^{2}={N_{0}\sum_{l}{a_{l}^{2}}}/{c^{2}}$.
Different choices of coefficients $a_{-\infty}^{\infty}$ provide
different bounds for $\Iach$. One appealing choice is the taps of
the sample whitened matched filter (SWMF), for which $\alpha_{k}=0$
for every $k\geq1$ \cite{lee2004digital}. This choice yields the
Shamai-Ozarow-Wyner bound \cite{shamai1991sow}:
\begin{equation}
\Iach\geq\Isow\triangleq I_{x}(\SNR[ZF-DFE])\label{eq:SOW bound}
\end{equation}
where
\begin{equation}
I_{x}(\gamma)\triangleq I\left(x_{0}\,;\,\sqrt{\gamma N_{0}/P_{x}}x_{0}+n_{0}\right)\label{eq:Ix_def}
\end{equation}
is the input-output mutual information in a scalar additive Gaussian
noise channel at SNR $\gamma$ and input distributed as a single ISI
channel input. $\SNR[ZF-DFE]$ stands for the output SNR of the unbiased
zero-forcing decision feedback equalizer (ZF-DFE), which uses the
SWMF as its front-end filter \cite{cioffi1995mmse}, and is given
by
\begin{equation}
\SNR[ZF-DFE]=\frac{P_{x}}{N_{0}}\exp\left\{ \frac{1}{2\pi}\int_{-\pi}^{\pi}\log\left(\left|H(\theta)\right|^{2}\right)d\theta\right\}
\end{equation}

Since evaluation of $I_{x}(\cdot)$ and $\SNR[ZF-DFE]$ amounts to
simple one-dimensional integration, the Shamai-Ozarow-Wyner bound
can be easily computed and analyzed. However, it is known to be quite
loose in medium and low SNR's.

Another choice of coefficients are the taps of the mean-squared whitened
matched filter (MS-WMF), for which the variance of the noise term
$\sum_{k\geq1}{\alpha_{k}x_{k}}+m$ is minimized. The MS-WMF is used
as the front-end filter of the MMSE-DFE \cite{cioffi1995mmse}. Denoting
the minimizing coefficients by $\left\{ \hat{\alpha}\right\} $ and
their corresponding Gaussian noise term by \textbf{$\hat{m}$}, the
SNR at the output of the unbiased MMSE-DFE is given by,
\begin{equation}
\SNR[DFE-U]=\frac{Ex_{0}^{2}}{E\left(\sum_{k\geq1}{\hat{\alpha}_{k}x_{k}}+\hat{m}\right)^{2}}=\exp\left\{ \frac{1}{2\pi}\int_{-\pi}^{\pi}\log\left(1+\frac{P_{x}}{N_{0}}\left|H(\theta)\right|^{2}\right)d\theta\right\} -1
\end{equation}
and we denote the resulting bound by
\begin{equation}
\Iach\geq\Immse\triangleq I\left(x_{0}\,;\, x_{0}+\sum_{k\geq1}{\hat{\alpha}_{k}x_{k}}+\hat{m}\right)\label{eq:Immse-def}
\end{equation}

The bound $\Immse$ is still difficult to handle numerically or analytically
because of the high complexity of the variable $\sum_{k\geq1}{\hat{\alpha}_{k}x_{k}}$.
Several techniques for further bounding $\Immse$ were proposed, such
as those in \cite{shamai1996intersymbol} and more recently in \cite{jeong2012easily}.
However, none of those methods provide bounds that are both simple
and tight.

In \cite{shamai1996intersymbol} Shamai and Laroia conjectured that
$\Immse$ can be lower bounded by replacing the interfering inputs
$x_{1}^{k}$ with i.i.d. Gaussian variables of the same variance,
i.e
\begin{gather}
\Immse\geq I\left(x_{0}\,;\, x_{0}+\sum_{k\geq1}{\hat{\alpha}_{k}g_{k}}+\hat{m}\right)=I_{x}(\SNR[DFE-U])\triangleq\Isl\label{eq:SLC}
\end{gather}
where $g_{k}$ are i.i.d. Gaussian variables with variance $P_{x}$
and independent of $x_{0}$ and $\hat{m}$. The inequality (\ref{eq:SLC})
is known as the Shamai-Laroia conjecture (SLC). The expression $\Isl$
was empirically shown to be a very tight approximation for $\Iach$
in a large variety of SNR's and ISI coefficients. Since it is also
elegant and easy to compute, the conjectured bound has seen much use
despite remaining unproven --- cf. \cite{jeong2012easily,arnold2001information,arnold2006simulation,kavcic2003binary}.

In a recent paper \cite{abbe2012coordinate}, Abbe and Zheng disproved
a stronger version of the SLC, by applying a geometrical tool using
Hermite polynomials. This so-called ``strong SLC'' claims that (\ref{eq:SLC})
holds true for any choice of coefficients $\alpha_{1}^{\infty}$,
and not just the MMSE coefficients ${\hat{\alpha}}_{1}^{\infty}$.
The disproof in \cite{abbe2012coordinate} is achieved by constructing
a counterexample in which the interference is composed of a single
tap (i.e. $\sum_{k\geq1}{\alpha_{k}x_{k}}=\alpha x_{1}$) and the
input distribution is a carefully designed small perturbation of a
Gaussian law. In this setting, it is shown that there exist SNR's
and values of $\alpha$ in which the strong SLC fails. In order to
apply this counterexample to the original SLC, one has to construct
appropriate ISI coefficients and their matching MMSE-DFE, which is
not trivial. Moreover, such a counterexample would use a continuous
input distribution, leaving room to hypothesize that the SLC holds
for practical finite-alphabet inputs.

The aim of this paper is to provide new insights into the validity
of the SLC, as well as to provide new simple lower bounds for $\Immse$.
Information-Estimation relations \cite{guo2005mutual} and related
results \cite{guo2011estimation} are instrumental in all of our analytic
results, as they enable the derivation of novel bounds and asymptotic
expressions for mutual information.

We begin by disproving the original (``weak'') SLC, showing analytically
that is does not hold when the SNR is sufficiently low, under very
general settings. Our proof relies on the power series expansion of
the input-output mutual information in the additive Gaussian channel
\cite{guo2011estimation}. This result allows us to construct specific
counterexamples in which computations clearly demonstrate that the
SLC does not hold. Furthermore, it provides insight on what makes
the Shamai Laroia expression such a good approximation, to the point
where it was never before observed not to hold in low SNR's.

With the SLC $\Immse\geq\Isl$ disproven, we are led to consider the
weakened but still highly meaningful conjecture, that $\Isl$ lower
bounds the achievable rate itself, i.e. $\Iach\geq\Isl$. We provide
numerical results indicating that for sufficiently skewed binary inputs
$\Iach<\Isl$ for some SNR, disproving the weakened bound in its most
general form. Nonetheless, we prove that for any finite entropy input
distribution and any ISI channel, the bound $\Iach\geq\Isl$ holds
for sufficiently high SNR. This proof is carried out by showing that
$\Iach$ converges to the input entropy at a higher exponential rate
than $\Isl$.

Finally, new bounds for $\Immse$ are proven using Information-Estimation
techniques and bounds on MMSE estimation of scaled sums of i.i.d.
variables contaminated by additive Gaussian noise. A simple parametric
bound is developed, which parameters can either be straightforwardly
optimized numerically or set to constant values in order to produce
an even simpler, if sometimes less tight, expression. Numerical results
are reported, showing the bounds to be useful in low to medium SNRs,
and of comparable tightness to that of the bounds reported in \cite{jeong2012easily}.

The rest of this paper is organized as follows. Section \ref{sec:low-snr}
contains the disproof of the original SLC via low-SNR asymptotic analysis.
Section \ref{sec:counterexamples} presents counterexamples for the
original SLC as well as the weakened bound $\Iach\geq\Isl$. Section
\ref{sec:high-snr} details the proof of the bound $\Iach\geq\Isl$
in the high-SNR regime, and section \ref{sec:bounds} established
novel Infromation-Estimation based bounds on $\Immse$. Section \ref{sec:conc}
concludes this paper.

\section{\label{sec:low-snr}Low SNR analysis of the Shamai-Laroia approximation}

In this section we prove that the conjectured bound (\ref{eq:SLC})
does not hold in the low SNR limit in essentially every scenario.
Given a zero-mean RV $x$, let $s_{x}=Ex^{3}/(Ex^{2})^{3/2}$ and
$\kappa_{x}=Ex^{4}/(Ex^{2})^{2}-3$ stand for its skewness and excess
kurtosis, respectively. Note that $s_{x}=\kappa_{x}=0$ for a Gaussian
RV. Our result is formally stated as,
\begin{thm}
\label{thm:slc_wrong}For every real ISI channel and any input with
$s_{x}=0$ and $\kappa_{x}\neq0$, \textup{$\Immse<\Isl$ }when \textup{$P_{x}/N_{0}$}
is sufficiently small. When $s_{x}\neq0$ there exist real ISI channels
for which \textup{$\Immse<\Isl$ }when \textup{$P_{x}/N_{0}$} is
sufficiently small.\end{thm}
\begin{IEEEproof}
The proof comprises of rewriting $\Immse-\Isl$ as a combination of
mutual informations in additive Gaussian channels, applying a fourth
order Taylor-series expansion to each element, and showing that the
resulting combination is always negative in the leading order.

First, let us state the Taylor expansion of the mutual information
in a useful form. Suppose $\xi$ is a zero-mean random variable and
let $\nu\sim\mathcal{N}(0,\sigma_{\nu}^{2})$ be independent of $\xi$.
It follows from equation (61) of \cite{guo2011estimation} that,

\begin{equation}
I\left(\xi\:;\:\xi+\nu\right)=\frac{\rho}{2}-\frac{\rho^{2}}{4}+\frac{\rho^{3}}{6}\left[1-\frac{s_{\xi}^{2}}{2}\right]-\frac{\rho^{4}}{48}\left[\kappa_{\xi}^{2}-12s_{\xi}^{2}+6\right]+O\left(\rho^{5}\right)\label{eq:Iexpansion}
\end{equation}

Where $\rho=E\xi^{2}/E\nu^{2}$. Let $\hat{\alpha}_{k}$ and $\hat{m}$
be the ISI coefficients and Gaussian noise term resulting from the
application of the unbiased MMSE-DFE filter on the channel output,
as defined in (\ref{eq:Immse-def}). In our proof we will make use
of the following definitions for $i=0,1$,

\begin{gather}
\mu_{i}\triangleq\sum_{k\geq i}\hat{\alpha}_{k}x_{k}\label{eq:mu01_def}\\
\tilde{\mu}_{1}\sim\mathcal{N}(0,E\mu_{1}^{2})\;,\ \tilde{\mu}_{1}\perp x_{0},\hat{m}\\
\tilde{\mu}_{0}\triangleq x_{0}+\tilde{\mu}_{1}\\
\beta_{i}^{2}\triangleq\sum_{k\geq i}{\hat{\alpha}_{k}}^{2}\;,\;\gamma_{i}^{3}\triangleq\sum_{k\geq i}{\hat{\alpha}_{k}}^{3}\;,\;\delta_{i}^{4}\triangleq\sum_{k\geq i}{\hat{\alpha}_{k}}^{4}\label{eq:beta01_def}\\
\epsilon_{i}=\frac{E\mu_{i}^{2}}{E{\hat{m}}^{2}}=\frac{E\tilde{\mu}_{i}^{2}}{E{\hat{m}}^{2}}=\frac{\beta_{i}^{2}P_{x}}{E{\hat{m}}^{2}}\label{eq:epsilon01_def}\\
\Immse^{(i)}\triangleq I\left(\mu_{i}\:;\:\mu_{i}+\hat{m}\right)\\
\Isl^{(i)}\triangleq I\left(\tilde{\mu}_{i}\:;\:\tilde{\mu}_{i}+\hat{m}\right)
\end{gather}
where $\hat{\alpha}_{0}\equiv1$. It is seen that
\begin{align*}
\Immse^{(0)} & =I\left(x_{0}+\mu_{1}\:;\: x_{0}+\mu_{1}+\hat{m}\right)=I\left(x_{0},x_{0}+\mu_{1}\:;\: x_{0}+\mu_{1}+\hat{m}\right)\\
 & =I\left(x_{0}\:;\: x_{0}+\mu_{1}+\hat{m}\right)+I\left(x_{0}+\mu_{1}\:;\: x_{0}+\mu_{1}+\hat{m}\:|\: x_{0}\right)\\
 & =\Immse+\Immse^{(1)}
\end{align*}
and so $\Immse=\Immse^{(0)}-\Immse^{(1)}$. Similarly, $\Isl=\Isl^{(0)}-\Isl^{(1)}$.
Let $\Delta_{i}\triangleq\Immse^{(i)}-\Isl^{(i)}$, it follows that,

\begin{equation}
\Immse-\Isl=\Delta_{0}-\Delta_{1}\label{eq:I_slc_diff_delta}
\end{equation}

Notice that $\Immse^{(0)},\Immse^{(1)},\Isl^{(0)},\Isl^{(1)}$ are
each the mutual information between the input and output of an additive
Gaussian channel, and can therefore readily be expanded according
to (\ref{eq:Iexpansion}), yielding
\begin{gather}
\Delta_{0}=-\left(\frac{\epsilon_{0}^{3}}{12}-\frac{\epsilon_{0}^{4}}{4}\right)\left[s_{\mu_{0}}^{2}-s_{\tilde{\mu}_{0}}^{2}\right]-\frac{\epsilon_{0}^{4}}{48}\left[\kappa_{\mu_{0}}^{2}-\kappa_{\tilde{\mu}_{0}}^{2}\right]+O\left(\epsilon_{0}^{5}\right)\label{eq:delta_0_comp}\\
\Delta_{1}=-\left(\frac{\epsilon_{1}^{3}}{12}-\frac{\epsilon_{1}^{4}}{4}\right)s_{\mu_{1}}^{2}-\frac{\epsilon_{1}^{4}}{48}\kappa_{\mu_{1}}^{2}+O\left(\epsilon_{1}^{5}\right)\label{eq:delta_1_comp}
\end{gather}
Where $s_{\tilde{\mu}_{1}}=\kappa_{\tilde{\mu}_{1}}=0$ since $\tilde{\mu}_{1}$
is Gaussian, and
\begin{alignat}{1}
s_{\mu_{i}} & =\frac{E\left(\sum_{k\geq i}\hat{\alpha}_{k}x_{k}\right)^{3}}{\beta_{i}^{3}P_{x}^{3/2}}=\frac{\gamma_{i}^{3}}{\beta_{i}^{3}}s_{x}\label{eq:skew_mu_i}\\
s_{\tilde{\mu}_{0}} & =\frac{E\left(x_{0}+\tilde{\mu}_{1}\right)^{3}}{\beta_{0}^{3}P_{x}^{3/2}}=\frac{s_{x}}{\beta_{0}^{3}}\label{eq:skew_x+g}\\
\kappa_{\mu_{i}} & =\frac{E\left(\sum_{k\geq i}\hat{\alpha}_{k}x_{k}\right)^{4}}{\beta_{i}^{4}P_{x}^{2}}-3=\frac{\delta_{i}^{4}}{\beta_{i}^{4}}\kappa_{x}\label{eq:kappa_mu_i}\\
\kappa_{\tilde{\mu}_{0}} & =\frac{E\left(x_{0}+\tilde{\mu}_{1}\right)^{4}}{\left(P_{x}+\beta_{1}^{2}P_{x}\right)^{2}}-3=\frac{\kappa_{x}}{\beta_{0}^{4}}\label{eq:kappa_x+g}
\end{alignat}
Putting everything together, we get:
\begin{equation}
\Immse-\Isl=-\frac{\gamma_{1}^{3}s_{x}^{2}}{6\beta_{0}^{6}}\epsilon_{0}^{3}-\left(\frac{\delta_{1}^{4}\kappa_{x}^{2}}{24\beta_{0}^{8}}-\frac{\left(2\beta_{0}^{2}+\gamma_{1}^{3}\right)\gamma_{1}^{3}s_{x}^{2}}{4\beta_{0}^{8}}\right)\epsilon_{0}^{4}+O\left(\epsilon_{0}^{5}\right)\label{eq:I_slc_diff_approx}
\end{equation}

For the case $s_{x}=0$,(\ref{eq:I_slc_diff_approx}) simplifies to,
\begin{equation}
\Immse-\Isl=-\frac{\delta_{1}^{4}\kappa_{x}^{2}}{24\beta_{0}^{8}}\epsilon_{0}^{4}+O\left(\epsilon_{0}^{5}\right)\label{eq:I_slc_diff_approx_zero_skew}
\end{equation}
=and clearly when $\epsilon_{0}\rightarrow0$ we must have $\Immse<\Isl$
from some point.

We now show that $\epsilon_{0}\rightarrow0$ when $P_{x}/N_{0}\rightarrow0$.
In Appendix \ref{app:second-order-expressions} we find that,
\begin{equation}
\epsilon_{0}=\SNR[LE]\frac{\SNR[DFE]-1}{\SNR[LE]-1}-1
\end{equation}
where $\SNR[LE]$, $\SNR[DFE]$ stand for the output SNR's of the
MMSE (biased) linear and decision-feedback equalizers, respectively
(see (\ref{eq:LE SNR def}) and (\ref{eq:DFE SNR def})). When $P_{x}/N_{0}$
is small, we have
\begin{gather}
\SNR[DFE]=\SNR[LE]=1+\left[\frac{1}{2\pi}{\displaystyle \intop_{-\pi}^{\pi}\left|H(\theta)\right|^{2}d\theta}\right]\frac{P_{x}}{N_{0}}+O\left(\left(\frac{P_{x}}{N_{0}}\right)^{2}\right)
\end{gather}
and therefore,
\begin{equation}
\epsilon_{0}=\left[\frac{1}{2\pi}{\displaystyle \intop_{-\pi}^{\pi}\left|H(\theta)\right|^{2}d\theta}\right]\frac{P_{x}}{N_{0}}+O\left(\left(\frac{P_{x}}{N_{0}}\right)^{2}\right)\label{eq:epsilon0_as_PxN0}
\end{equation}
and goes to zero when $P_{x}/N_{0}\rightarrow0$. This proves our
statement in the case $s_{x}=0$, since by (\ref{eq:I_slc_diff_approx_zero_skew})
and (\ref{eq:epsilon0_as_PxN0}), the leading term in the expansion
of $\Immse-\Isl$ with respect to $P_{x}/N_{0}$ is guaranteed to
be negative.

When $s_{x}\neq0$, we will demonstrate that there exist ISI channels
for which $\gamma_{1}>0$ at low SNRs. Let us consider the two tap
channel $h\left(D\right)=\sqrt{1-q^{2}}+qD^{-1}$ for some $0<q<1$.
Carrying out the calculation according to \cite{cioffi1995mmse} reveals
that the residual ISI satisfies,
\begin{equation}
\alpha_{i}^{*}=\frac{\left(-1\right)^{i+1}\left[a-\sqrt{a^{2}-1}\right]^{i}}{\frac{1}{2}\left[1+\sqrt{1-1/a^{2}}\right]\left(1+\frac{P_{x}}{N_{0}}\right)-1}
\end{equation}
where
\begin{equation}
a=\frac{1+N_{0}/P_{x}}{2q\sqrt{1-q^{2}}}\geq1
\end{equation}
Thus, for small $P_{x}/N_{0}$ one finds that
\begin{equation}
\gamma_{1}^{3}=q^{3}\left(1-q^{2}\right)^{3/2}+O\left(\frac{P_{x}}{N_{0}}\right)\label{eq:gamma3 binary channel}
\end{equation}
Plugging (\ref{eq:gamma3 binary channel}) into (\ref{eq:I_slc_diff_approx})
and (\ref{eq:epsilon0_as_PxN0}), we conclude that for channels of
the form $h\left(D\right)=\sqrt{1-q^{2}}+qD^{-1}$ with $0<q<1$,
\begin{equation}
\Immse-\Isl=-\frac{1}{6}q^{3}\left(1-q^{2}\right)^{3/2}s_{x}^{2}\left(\frac{P_{x}}{N_{0}}\right)^{3}+O\left(\left(\frac{P_{x}}{N_{0}}\right)^{4}\right)
\end{equation}
proving our statement for the case of non-zero skewness.
\end{IEEEproof}

\section{\label{sec:counterexamples}Counterexamples}

In this section we use insights from Section \ref{sec:low-snr} in
order to construct specific counterexamples for the SLC in both its
original= form ($\Immse\geq\Isl$) and its weakened version ($\Iach\geq\Isl$).
The section is composed of two parts. In the first part we compare
$\Immse$ and $\Isl$ in the low-SNR regime for specific input distributions
and ISI channel, demonstrating Theorem \ref{thm:slc_wrong} and verifying
the series expansion derived in its proof. In the second part we compare
$\Iach$ and $\Isl$, with the former estimated by means of Monte-Carlo
simulation, in the medium-SNR regime and with the ISI channel and
input distributions that were used in the first part of this section.
Our results indicate that for highly skewed binary inputs, $\Iach<\Isl$
for some SNRs.

\subsection{Low-SNR regime --- $\Immse<\Isl$}

\begin{figure}
\begin{centering}
\includegraphics[width=14cm]{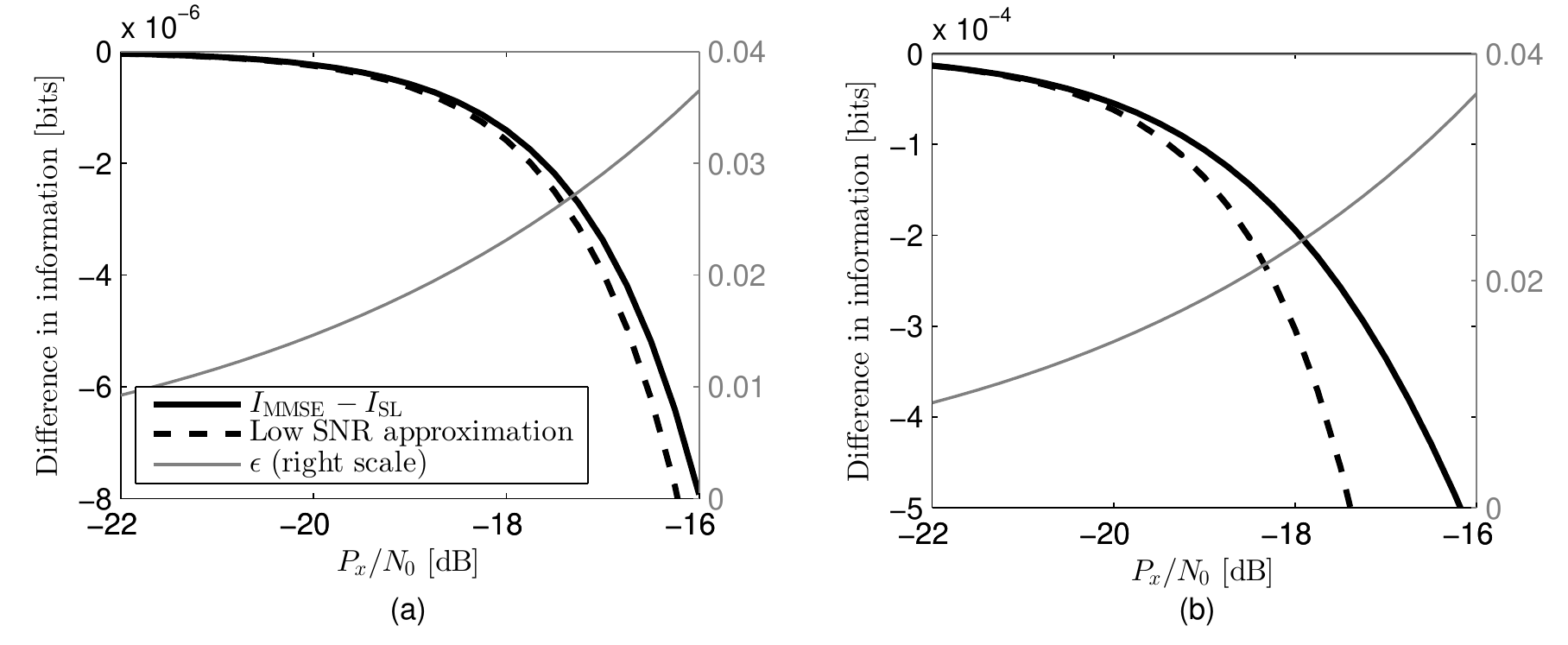}
\par\end{centering}

\caption{\label{fig:low-snr-diff}$\Immse-\Isl$ and $\epsilon_{0}$ for (a)
trinary input with high kurtosis and (b) highly skewed binary input,
in the low SNR regime and under moderate ISI.}
\end{figure}

Figure \ref{fig:low-snr-diff} demonstrates Theorem \ref{thm:slc_wrong}
and its inner workings, for a particular choice of ISI coefficients
and two input distributions. The first distribution represents a symmetric
source with input alphabet $\left\{ -1,0,1\right\} $ and $\Pr\left(x=1\right)=0.01$,
that has zero skewness and excess kurtosis $\kappa_{x}=47$. The second
distribution represents a zero-mean skewed binary source with $\Pr\left(X>0\right)=0.002$,
that has $s_{x}\approx-22.3$ and $\kappa_{x}\approx495$. The ISI
is formed by a three taps impulse response with $h_{0}=h_{2}=0.408$
and $h_{1}=0.817$ (``Channel B'' from \cite{proakis1987digital}
chapter 10). Examining Figure \ref{fig:low-snr-diff} it is seen that
in the low SNR regimem $\Immse-\Isl$ is indeed negative and well
approximated by the expansion (\ref{eq:I_slc_diff_approx}) --- in
agreement with Theorem \ref{thm:slc_wrong} and in contradiction to
the Shamai-Laroia conjecture.

In order to estimate $\Immse$ as defined in (\ref{eq:Immse-def}),
the infinite sequence of residual ISI taps ${\hat{\alpha}}_{1}^{\infty}$
is truncated to ${\hat{\alpha}}_{1}^{N}$ with the minimal $N$ for
which $\sum_{k>N}{\hat{\alpha}_{k}}^{2}<10^{-10}\sum_{k\geq1}{\hat{\alpha}_{k}}^{2}$
. For the ISI channel used in our counterexample, $N$ moves from
8 at SNR -26 dB to 36 at SNR 10 dB. Experimentation indicates that
the accuracy of the computation of $\Immse$ and $\Isl$ is of the
order of $10^{-9}$ bit.

To clearly observe the behavior predicted by Theorem \ref{thm:slc_wrong},
it is crucial to use an input distribution with high skewness or high
kurtosis. Using the notation of (\ref{eq:I_slc_diff_approx}), we
observe that the difference $\Immse-\Isl$ is of the order of $s_{x}^{2}\epsilon_{0}^{3}\gamma_{1}^{3}/\beta_{0}^{6}+\kappa_{x}^{2}\epsilon_{0}^{4}\delta_{1}^{4}/\beta_{0}^{8}$.
Computations reveal that for channels with moderate to high ISI, $\left|\gamma_{1}^{3}\right|/\beta_{0}^{6}$
and $\delta_{1}^{4}/\beta_{0}^{8}$ are both of the order of $0.05$
at low SNRs, and that the series approximation is valid up to $\epsilon_{0}$
values of around $0.02$. Hence, the difference term is roughly $10^{-8}s_{x}^{2}+10^{-10}\kappa_{x}^{2}$.
Therefore, we must have $s_{x}^{2}$ of the order of $10$ and/or
$\kappa_{x}^{2}$ of the order of $10^{3}$ for the predicted low-SNR
behavior to be distinguishable from numeric errors.

We emphasize that Theorem 1 guarantees that the SLC does not hold
for any input distribution with nonzero skewness or excess kurtosis,
including for example BPSK input that has $s_{x}=0$ and $\kappa_{x}=-2$.
However, the above analysis shows that the universal low SNR behavior
(\ref{eq:I_slc_diff_approx}) is masked by numerical errors when common
input distributions are used, due to the facts that by symmetry they
have zero skewness, and that their excess kurtosis values are of order
unity. This serves to explain why similar low SNR counterexamples
to the SLC were not previously reported.

\subsection{Medium-SNR regime --- $\Iach<\Isl$}

\begin{figure}
\centering{}\includegraphics[width=16cm]{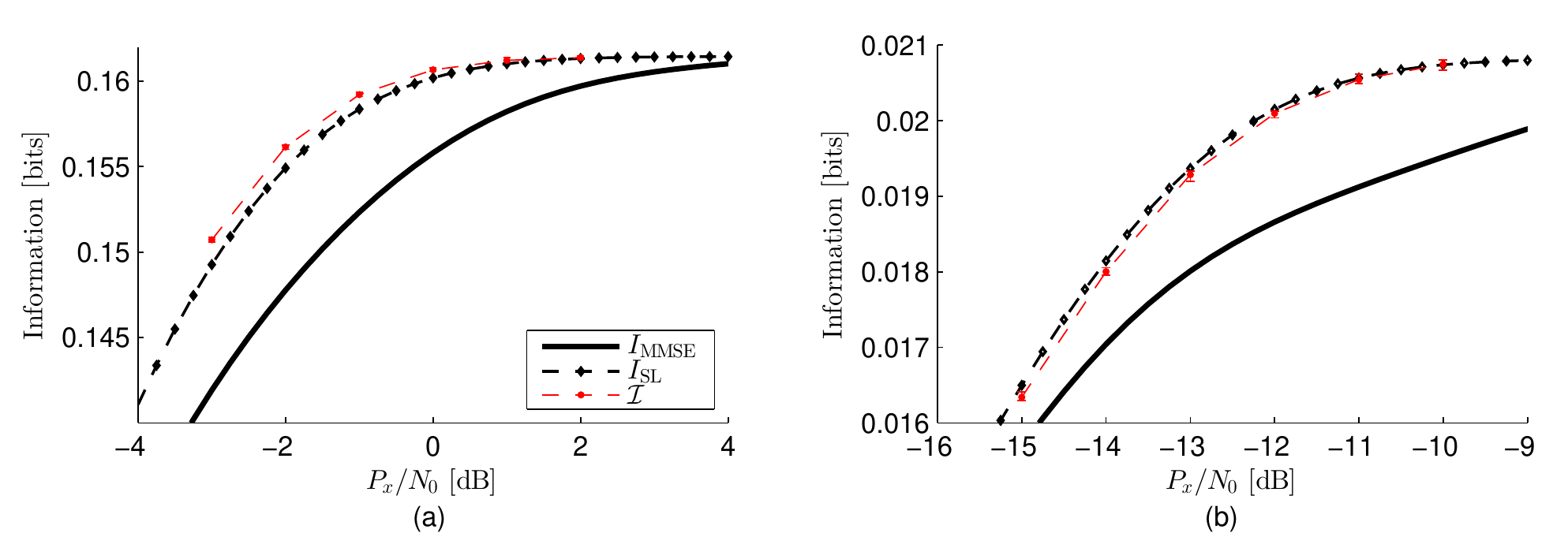}\caption{\label{fig:high_kurt_with_mc}$\Immse$, $\Isl$ and $\Iach$ for
(a) trinary input with high kurtosis and (b) highly skewed binary
input, in the medium SNR regime and under moderate ISI.}
\end{figure}

Figure \ref{fig:high_kurt_with_mc} displays $\Immse$, $\Isl$ and
$\Iach$ computed for the input distributions and ISI channel described
above. The value of $\Iach$ is computed by Monte-Carlo simulations
as described in \cite{arnold2006simulation}. For each SNR, 20 simulations
with input length $5\cdot10^{8}$ were preformed. The dots on the
red curve indicate the averaged result of these simulations (which
is equivalent to a single simulation with input length $10^{10}$),
and the error bars indicate the minimum and maximum results among
the 20 simulations.

For both input distributions, $\Isl$ clearly exceeds $\Immse$. In
fact, further simulations indicate that in both cases $\Isl>\Immse$
for the entire SNR range, leaving little room to hope that the Shamai-Laroia
conjecture is valid in the high-SNR regime. For the symmetric trinary
source, it is seen that $\Iach>\Isl$ for all SNRs tested. However,
for the skewed binary sources, it is fairly certain that $\Iach<\Isl$
at some SNRs. This leads to the conclusion that even the modified
conjecture $\Iach\geq\Isl$ does not hold in general.

The relation $\Iach\geq\Isl$ might still be true for all SNRs and
ISI channels for some input distributions, such as BPSK, and might
even hold for large families of input distributions, such as symmetric
sources. Our simulations indicate that $\Isl$ is always a tight approximation
for $\Iach$, and that it is much tighter than $\Immse$ for sources
with high skewness or excess kurtosis. Moreover, in the following
section we establish that in the high-SNR regime, the inequality $\Iach\geq\Isl$
holds for any input distribution and any ISI channel.

\section{\label{sec:high-snr}High SNR analysis of the Shamai-Laroia approximation}

In this section we prove that the weakened Shamai-Laroia bound $\Iach\geq\Isl$
is valid for any input distribution and ISI channel, for sufficiently
high SNR. The proof is carried out by bounding the exponential rates
at which $\Iach$ and $\Isl$ converge to the input entropy as the
input SNR grows, and showing that the former rate is strictly higher
than the latter for every non-trivial ISI channel. The rate of convergence
of $\Iach$ is lower bounded using Fano's inequality and Forney's
analysis of the probability of error of the Maximum Likelihood sequence
detector of the input to the ISI channel given its output. The rate
of convergence of $\Isl$ is upper bounded using the I-MMSE relationship
and genie-based bounds on the MMSE estimation of a single channel
input from an observation contaminated by additive Gaussian noise.

For convenience, the results of this section assume the normalization
$\sum_{k=0}^{L-1}h_{k}^{2}=\frac{1}{2\pi}\int_{-\pi}^{\pi}\left|H\left(\theta\right)\right|^{2}d\theta=1$.
Let
\begin{equation}
g_{\textrm{ZF-DFE}}=\exp\left\{ \frac{1}{2\pi}\int_{-\pi}^{\pi}\log\left(\left|H\left(\theta\right)\right|^{2}\right)d\theta\right\}
\end{equation}
denote the gain factor of the zero-forcing DFE --- It is seen that
$\SNR[DFE]$ behaves as $\frac{P_{x}}{N_{0}}g_{\textrm{ZF-DFE}}$
when $P_{x}/N_{0}\rightarrow\infty$. For every possible channel input
$x\in\mathcal{X}$, let $p\left(x\right)$ denote its probability
of occurrence and let $H\left(x_{0}\right)=-\sum_{x\in\mathcal{X}}p\left(x\right)\log p\left(x\right)$
be the input entropy. Finally, let $d_{\min}=\min_{x,x'\in\mathcal{X}}\left|x-x'\right|$
denote the minimal distance between different input values.

Our asymptotic bound for the achievable rate $\Iach$ is formally
stated as follows,
\begin{lem}
\label{lem:achievable rate high SNR bound}For any finite entropy
input distribution and any finite length ISI channel there exists
a function $F\left(x\right)>0$ polynomial in $x$ and a constant
$\delta_{\min}^{2}$such that,
\begin{equation}
H\left(x_{0}\right)-\Iach\leq F\left(\frac{P_{x}}{N_{0}}\right)\exp\left(-\frac{P_{x}}{2N_{0}}\left(\frac{d_{\min}}{2}\right)^{2}\delta_{\min}^{2}\right)\label{eq:C ML bound}
\end{equation}
and $\delta_{\min}^{2}\geq g_{\textrm{ZF-DFE}}$, with strict inequality
whenever $\left|H\left(\theta\right)\right|$ is not constant (i.e.
there is non-zero ISI).\end{lem}
\begin{IEEEproof}
Since $x_{-\infty}^{\infty}$ is i.i.d., $H\left(x_{0}|x_{-\infty}^{-1}\right)=H\left(x_{0}\right)$
and hence
\begin{equation}
H\left(x_{0}\right)-\Iach=H\left(x_{0}|y_{-\infty}^{\infty},x_{-\infty}^{-1}\right)\leq H\left(x_{0}|y_{-\infty}^{\infty}\right)=H\left(x_{0}|\hat{x}_{0}^{ML},y_{-\infty}^{\infty}\right)\leq H\left(x_{0}|\hat{x}_{0}^{ML}\right)\label{eq:ML inequality}
\end{equation}
where $\left\{ \hat{x}_{i}^{ML}\right\} _{i=-\infty}^{\infty}$ is
the maximum likelihood sequence estimate of $x_{-\infty}^{\infty}$
given $y_{-\infty}^{\infty}$. By Fano's inequality,
\begin{eqnarray}
H\left(x_{0}|\hat{x}_{0}^{ML}\right) & \leq & H\left(x_{0},1_{\left\{ x_{0}=\hat{x}_{0}^{ML}\right\} }|\hat{x}_{0}^{ML}\right)\leq H\left(1_{\left\{ x_{0}=\hat{x}_{0}^{ML}\right\} }\right)+H\left(x_{0}|1_{\left\{ x_{0}=\hat{x}_{0}^{ML}\right\} },\hat{x}_{0}^{ML}\right)\\
 & \leq & h_{2}\left(\Pr\left(x_{0}\neq\hat{x}_{0}^{ML}\right)\right)+\Pr\left(x_{0}\neq\hat{x}_{0}^{ML}\right)\log\left|\mathcal{X}\right|\label{eq:ML fano bound}
\end{eqnarray}
\begin{equation}
H\left(x_{0}|\hat{x}_{0}^{ML}\right)\leq h_{2}\left(\Pr\left(x_{0}\neq\hat{x}_{0}^{ML}\right)\right)+\Pr\left(x_{0}\neq\hat{x}_{0}^{ML}\right)\log\left|\mathcal{X}\right|
\end{equation}
where $h_{2}\left(x\right)=-x\log x-\left(1-x\right)\log\left(1-x\right)$
is the binary entropy function and $\mathcal{X}$ is the set of possible
values of $x_{0}$. By the analysis of the probability of error in
maximum likelihood sequence estimation first preformed by Forney \cite{forney1972maximum}
and then refined in \cite{foschini1975performance,wyner1975upper,verdu1987maximum},
we know that
\begin{equation}
\Pr\left(x_{0}\neq\hat{x}_{0}^{ML}\right)\leq K'Q\left(\sqrt{\frac{P_{x}}{N_{0}}\left(\frac{d_{\min}}{2}\right)^{2}\delta_{\min}^{2}}\right)\label{eq:ML perr}
\end{equation}
with $K'>0$ and $\delta_{\min}^{2}$ the minimum weighted and normalized
distance between any two input sequences that first diverge at time
$0$ and last diverge at some finite time $N$,
\begin{equation}
\delta_{\min}^{2}=\inf_{N\geq1}\min_{\begin{array}{c}
x_{0}^{N-1},\tilde{x}_{0}^{N-1}\mbox{ s.t. }\\
x_{0}\neq\tilde{x}_{0},x_{N-1}\neq\tilde{x}_{N-1}
\end{array}}\delta^{2}\left(x_{0}^{N-1},\tilde{x}_{0}^{N-1}\right)\label{eq:delta min def}
\end{equation}
where
\begin{equation}
\delta^{2}\left(x_{0}^{N-1},\tilde{x}_{0}^{N-1}\right)=\sum_{k=0}^{L+N-2}\left|\sum_{l=0}^{N-1}\left(\frac{x_{l}-\tilde{x}_{l}}{d_{\min}}\right)h_{k-l}\right|^{2}\label{eq:delta2 def}
\end{equation}

Substituting (\ref{eq:ML perr}) into (\ref{eq:ML fano bound}) and
taking (\ref{eq:ML inequality}) into account, along with the fact
that $Q\left(x\right)\leq\frac{1}{\sqrt{2\pi}}e^{-x^{2}/2}$, yields
the bound (\ref{eq:C ML bound}). It remains to show that $\delta_{\min}^{2}$
can be lower bounded by $g_{\textrm{ZF-DFE}}$. By keeping only the
first and last summands in (\ref{eq:delta2 def}), we have that when
$L>1$, for any feasible pair of sequences $x_{0}^{N-1},\tilde{x}_{0}^{N-1}$,
\begin{equation}
\delta^{2}\left(x_{0}^{N-1},\tilde{x}_{0}^{N-1}\right)\geq\left|\left(\frac{x_{0}-\tilde{x}_{0}}{d_{\min}}\right)h_{0}\right|^{2}+\left|\left(\frac{x_{N-1}-\tilde{x}_{N-1}}{d_{\min}}\right)h_{L-1}\right|^{2}\geq\left|h_{0}\right|^{2}+\left|h_{L-1}\right|^{2}
\end{equation}
since by assumption $x_{0}\neq\tilde{x}_{0}$ and $x_{N-1}\neq\tilde{x}_{N-1}$,
and $x\neq\tilde{x}$ implies $\left|x-\tilde{x}\right|\geq d_{\min}$.
Hence, $\delta_{\min}^{2}\geq\left|h_{0}\right|^{2}+\left|h_{L-1}\right|^{2}>\left|h_{0}\right|^{2}$
for $L>1$.

We may assume without loss of generality that $H\left(\theta\right)$
is minimum phase (i.e. has no zeros outside the unit circle), because
it may always be brought to this form by means of a whitened matched
filter. When $H\left(\theta\right)$ is minimum phase it follows that
$g_{\textrm{ZF-DFE}}=\left|h_{0}\right|^{2}$, and thus we conclude
that $\delta_{\min}^{2}>g_{\textrm{ZF-DFE}}$, except for the zero-ISI
case $L=1$. For $L=1$, $\delta_{\min}^{2}=g_{\textrm{ZF-DFE}}=1$.
\end{IEEEproof}
Our asymptotic bound for the achievable rate for the Shamai-Laroia
expression $\Isl$ is given as,
\begin{lem}
\label{lem:ISL high SNR bound}For any finite entropy input distribution
and any finite length ISI channel there exists a function $G\left(x\right)>0$
polynomial in $x$ and constants $\varepsilon,\hat{K}>0$ such that,
\begin{equation}
H\left(x_{0}\right)-\Isl\geq G\left(\frac{P_{x}}{N_{0}}\right)\exp\left(-\frac{P_{x}}{2N_{0}}\left(\frac{d_{\min}}{2}\right)^{2}g_{\textrm{ZF-DFE}}-\hat{K}\cdot\left(\frac{P_{x}}{N_{0}}\right)^{1-\varepsilon}\right)
\end{equation}
\end{lem}
\begin{IEEEproof}
We rewrite $I_{x}\left(\mathrm{snr}\right)$ as defined in (\ref{eq:Ix_def})
using the I-MMSE relation \cite{guo2005mutual},
\begin{equation}
H\left(x_{0}\right)-I_{x}\left(\mathrm{snr}\right)=\frac{1}{2}\int_{\mathrm{snr}}^{\infty}\mathrm{mmse}_{\bar{x}}\left(\gamma\right)d\gamma\label{eq:ISL IMMSE high SNR}
\end{equation}
where $\bar{x}=x/\sqrt{P_{x}}$, and for any RV $z$,
\begin{equation}
\mathrm{mmse}_{z}\left(\gamma\right)\triangleq E\left(z-E\left[z\,|\,\sqrt{\gamma}z+n\right]\right)^{2}\label{eq:mmsex-def}
\end{equation}
with $n\sim\mathcal{N}\left(0,1\right)$ and independent of $z$.
Let $v_{1}$ and $v_{2}$ be two possible values of $x$ such that
$\left|v_{1}-v_{2}\right|=d_{\min}$, and denote their probabilities
$p\left(v_{1}\right)$ and $p\left(v_{2}\right)$, respectively, assuming
without loss of generality that $p\left(v_{1}\right)\leq p\left(v_{2}\right)$.
Let $U$ be a random variable independent of $x$ and distributed
on $\left\{ 0,1\right\} $ with $\Pr\left(U=1\right)=p\left(v_{1}\right)/p\left(v_{2}\right)$.
Define the random variable $B=1_{\left\{ x=v_{1}\right\} }+U\cdot1_{\left\{ x=v_{2}\right\} }$,
so that given $B=1$, $x$ is distributed equiprobably on $\left\{ v_{1},v_{2}\right\} $.
Since conditioning can only decrease MMSE we have
\begin{equation}
\mathrm{mmse}_{\bar{x}}\left(\gamma\right)\geq\mathrm{mmse}_{\bar{x}|B}\left(\gamma\right)\geq\Pr\left(B=1\right)\mathrm{mmse}_{\bar{x}|B=1}\left(\gamma\right)\label{eq:MMSE genie bound non-equiprobable}
\end{equation}

Now, $\Pr\left(B=1\right)=2p\left(v_{1}\right)$, and $\mathrm{mmse}_{\bar{x}|B=1}\left(\gamma\right)=\left(\frac{d_{\min}}{2}\right)^{2}\mathrm{mmse}_{b}\left(\left(\frac{d_{\min}}{2}\right)^{2}\gamma\right)$,
where $b$ is equiprobably distributed on $\left\{ -1,1\right\} $.
The function $\mathrm{mmse}_{b}$ can be bounded as
\begin{eqnarray}
\mathrm{mmse}_{b}\left(\gamma\right) & = & \frac{1}{\sqrt{2\pi}}\int_{-\infty}^{\infty}\left(1-\tanh\left(\sqrt{\gamma}y\right)\right)e^{-\frac{1}{2}\left(y-\sqrt{\gamma}\right)^{2}}dy\\
 & \geq & \frac{1}{\sqrt{2\pi}}\int_{-\infty}^{\infty}e^{-\sqrt{\gamma}\left(y+\left|y\right|\right)}e^{-\frac{1}{2}\left(y-\sqrt{\gamma}\right)^{2}}=2Q\left(\sqrt{\gamma}\right)
\end{eqnarray}
where we have used $1-\tanh\left(x\right)\geq e^{-x-\left|x\right|}$.
Using $Q\left(\sqrt{x}\right)<\frac{e^{-x/2}}{\sqrt{2\pi x}}$ we
find that,
\begin{equation}
\int_{s}^{\infty}Q\left(\sqrt{\gamma}\right)d\gamma=\frac{\sqrt{s}e^{-s/2}}{\sqrt{2\pi}}+\left(1-s\right)Q\left(\sqrt{s}\right)\underset{s>1}{\geq}\frac{e^{-s/2}}{\sqrt{2\pi s}}
\end{equation}
and so
\begin{equation}
\frac{1}{2}\int_{\mathrm{snr}}^{\infty}\mathrm{mmse}_{\bar{x}}\left(\gamma\right)d\gamma\geq p\left(v_{1}\right)\int_{\left(\frac{d_{\min}}{2}\right)^{2}\mathrm{snr}}^{\infty}\mathrm{mmse}_{b}\left(\gamma\right)d\gamma\geq\frac{2\sqrt{2}p\left(v_{1}\right)}{\sqrt{\pi d_{\min}^{2}\mathrm{snr}}}\exp\left(-\frac{1}{2}\mathrm{\left(\frac{d_{\min}}{2}\right)^{2}snr}\right)\label{eq:MMSE lower bound high SNR}
\end{equation}
We remark that a bound similar to (\ref{eq:MMSE genie bound non-equiprobable})
was developed in \cite{lozano2006optimum}. However, the lower bound
of \cite{lozano2006optimum} does not take into account non-equiprobable
inputs.

The last step is to upper bound $\SNR[DFE]$ in terms of $g_{\textrm{ZF-DFE}}$
for large $P_{x}/N_{0}$. We have
\begin{equation}
\log\left(\frac{\SNR[DFE]}{\frac{P_{x}}{N_{0}}g_{\textrm{ZF-DFE}}}\right)=\frac{1}{2\pi}\int_{-\pi}^{\pi}\log\left(1+\frac{N_{0}}{P_{x}\left|H\left(\theta\right)\right|^{2}}\right)d\theta\label{eq:MS ZF log ratio}
\end{equation}
If $\left|H\left(\theta\right)\right|^{2}>0$ for every $\theta$,
a simple bound is obtained using $\frac{1}{2\pi}\int_{-\pi}^{\pi}\log\left(f\left(\theta\right)\right)d\theta\leq\log\left(\frac{1}{2\pi}\int_{-\pi}^{\pi}f\left(\theta\right)d\theta\right)$:
\begin{equation}
\SNR[DFE]\leq\frac{P_{x}}{N_{0}}g_{\textrm{ZF-DFE}}+\frac{g_{\textrm{ZF-DFE}}}{g_{\textrm{ZF-LE}}}
\end{equation}
with $g_{\textrm{ZF-LE}}=\left[\frac{1}{2\pi}\int_{-\pi}^{\pi}\frac{d\theta}{\left|H\left(\theta\right)\right|^{2}}\right]^{-1}$
being the SNR gain factor of the linear zero-forcing equalizer. However,
if the channel has spectral nulls, $g_{\textrm{ZF-LE}}=0$ and the
above bound is useless. In this case, let
\begin{equation}
\Omega=\left\{ \theta\in\left[-\pi,\pi\right]\:|\:\left|H\left(\theta\right)\right|^{2}<\sqrt{N_{0}/P_{x}}\right\}
\end{equation}
and for $N_{0}/P_{x}<1$, bound (\ref{eq:MS ZF log ratio}) as
\begin{equation}
\frac{1}{2\pi}\int_{-\pi}^{\pi}\log\left(1+\frac{N_{0}}{P_{x}\left|H\left(\theta\right)\right|^{2}}\right)d\theta\leq\frac{1}{2\pi}\int_{\Omega}\log\left(2\sqrt{\frac{N_{0}}{P_{x}}}\frac{1}{\left|H\left(\theta\right)\right|^{2}}\right)d\theta+\frac{\left|\Omega^{C}\right|}{2\pi}\log\left(1+\sqrt{\frac{N_{0}}{P_{x}}}\right)
\end{equation}
The second term above is upper bounded simply as $\log\left(1+\sqrt{\frac{N_{0}}{P_{x}}}\right)$,
while the first term is bounded using the Cauchy-Schwartz inequality
assuming now $2\sqrt{N_{0}/P_{x}}<1$:
\begin{alignat}{1}
\frac{1}{2\pi}\int_{\Omega}\log\left(2\sqrt{\frac{N_{0}}{P_{x}}}\frac{1}{\left|H\left(\theta\right)\right|^{2}}\right)d\theta & \leq\frac{1}{2\pi}\int_{-\pi}^{\pi}1_{\left\{ \theta\in\Omega\right\} }\log\left(\frac{1}{\left|H\left(\theta\right)\right|^{2}}\right)d\theta\\
\leq & \sqrt{\frac{\left|\Omega\right|}{2\pi}}\sqrt{\frac{1}{2\pi}\int_{-\pi}^{\pi}\log^{2}\left(\left|H\left(\theta\right)\right|^{2}\right)d\theta}
\end{alignat}
Since $\left|H\left(\theta\right)\right|^{2}=c\prod_{i=1}^{L-1}\left(1-\zeta_{i}\cos\left(\theta-\theta_{i}\right)\right)$
with $c>0$, $\left|\zeta_{i}\right|\leq1$ and $\theta_{i}\in\left[-\pi,\pi\right]$
it can be shown that $\frac{1}{2\pi}\int_{-\pi}^{\pi}\log^{2}\left(\left|H\left(\theta\right)\right|^{2}\right)d\theta<c_{1}^{2}<\infty$
and $\left|\Omega\right|<2\pi c_{2}\left(N_{0}/P_{x}\right)^{\varepsilon'}$
for some $c_{1},c_{2,},\varepsilon'>0$. Therefore,
\begin{equation}
\SNR[DFE]\leq\frac{P_{x}}{N_{0}}g_{\textrm{ZF-DFE}}\left(1+\sqrt{\frac{N_{0}}{P_{x}}}\right)e^{c_{1}c_{2}\left(N_{0}/P_{x}\right)^{\varepsilon'}}\leq\frac{P_{x}}{N_{0}}g_{\textrm{ZF-DFE}}+\hat{K}\left(\frac{P_{x}}{N_{0}}\right)^{1-\varepsilon}\label{eq:SNR MS DFE bound}
\end{equation}
for some $\varepsilon,\hat{K}>0$ and sufficiently large $P_{x}/N_{0}$.

Using (\ref{eq:SLC}), (\ref{eq:ISL IMMSE high SNR}), (\ref{eq:MMSE lower bound high SNR})
and (\ref{eq:SNR MS DFE bound}) we conclude that
\begin{alignat}{1}
H\left(x_{0}\right)-\Isl & \geq H\left(x_{0}\right)-I_{x}\left(\frac{P_{x}}{N_{0}}g_{\textrm{ZF-DFE}}+\hat{K}\cdot\left(\frac{P_{x}}{N_{0}}\right)^{1-\varepsilon}\right)\nonumber \\
\geq & G\left(\frac{P_{x}}{N_{0}}\right)\exp\left(-\frac{P_{x}}{2N_{0}}\left(\frac{d_{\min}}{2}\right)^{2}g_{\textrm{ZF-DFE}}-\hat{K}\cdot\left(\frac{P_{x}}{N_{0}}\right)^{1-\varepsilon}\right)
\end{alignat}
with $G\left(x\right)$ given by $2\sqrt{2}p\left(v_{1}\right)\left[\pi d_{\min}^{2}\left(g_{\textrm{ZF-DFE}}x+\hat{K}x^{1-\varepsilon}\right)\right]^{-1/2}$
.
\end{IEEEproof}
Using the above results, we are able to state our desired conclusion,
\begin{thm}
\label{thm:SLC proof high snr}For any finite entropy input distribution
and any finite length ISI channel, $\Iach\geq\Isl$ for sufficiently
high SNR.\end{thm}
\begin{IEEEproof}
Immediate from Lemmas \ref{lem:achievable rate high SNR bound} and
\ref{lem:ISL high SNR bound}.
\end{IEEEproof}

\section{\label{sec:bounds}Information-Estimation-based bounds for $\Immse$}

Having shown that $\Isl$ is not always a lower bound on $\Immse$
and that sometimes it is not even a lower bound for $\Iach$, in this
section we establish new lower bounds for $\Immse$ and hence for
$\Iach$. The bounds are based on a simple genie-based bound for the
MMSE in estimating a linear combination of i.i.d. variables from their
Gaussian noise corrupted version, that are related to $\Immse$ via
the Guo-Shamai-Verdú theorem. A general bound with two scalar parameters
is derived. These parameters may be either easily optimized numerically,
or fixed in order to yield a simpler expression, which is optimal
for low SNR's and nearly as tight in the high SNR regime. The bounds
are evaluated and compared with recently proposed lower bounds for
$\Immse$ where the input is binary. They are found to be quite tight
at low SNR's, reasonable at medium SNR's and loose for high SNR's.

\subsection{A general MMSE bound}

Consider the random variable

\begin{equation}
X=\sum_{k\in\Kset}a_{k}x_{k}
\end{equation}
where $x_{k}$ are i.i.d. RVs with $Ex^{2}=1$, $\Kset\subseteq\mathbb{N}$
and the coefficients $\left\{ a_{k}\right\} _{k\in\Kset}$ satisfy
$\sum_{k\in\Kset}a_{k}^{2}=1$. Let $Y=\sqrt{\gamma}X+N$ with $N$
a standard Gaussian variable independent of $X$, so that by (\ref{eq:mmsex-def}),
\begin{equation}
\textrm{mmse}_{X}(\gamma)=E\left(X-E[X\,|\, Y]\right)^{2}
\end{equation}

Let $\left\{ \Pset_{m}\right\} _{m\in\Mset}$ be a partition of $\Kset$
and define for every $m\in\Mset$,
\begin{eqnarray}
b_{m}^{2} & = & \sum_{k\in\Pset_{m}}a_{k}^{2}\\
X_{m} & =\frac{1}{b_{m}} & \sum_{k\in\Pset_{m}}a_{k}x_{k}
\end{eqnarray}
and
\begin{equation}
Y_{m}=\sqrt{\gamma}X_{m}+N_{m}
\end{equation}
where $\left\{ N_{m}\right\} _{m\in\Mset}$ are independent with $N_{m}\sim\mathcal{N}(0,\sigma_{m}^{2})$
and satisfy $\sum_{m\in\Mset}b_{m}^{2}\sigma_{m}^{2}=1$.
\begin{lem}
\label{lem:mmse_iid_sum_bound}Under the above definitions,
\begin{equation}
\textrm{mmse}_{X}(\gamma)\geq\sum_{m\in\Mset}b_{m}^{2}\textrm{mmse}_{X_{m}}(\frac{\gamma}{\sigma_{m}^{2}})\label{eq:mmse-genie-general}
\end{equation}
\end{lem}
\begin{IEEEproof}
Note that we may write $Y=\sum_{m\in\Mset}b_{m}Y_{m}$. Since conditioning
decreases MMSE,
\begin{equation}
\textrm{mmse}_{X}(\gamma)\geq E\left(X-E\left[X|\left\{ Y_{m}\right\} _{m\in\Mset}\right]\right)^{2}
\end{equation}
For every $m\neq m'$, $X_{m}$ is independent of $X_{m'}$ and $Y_{m'}$.
Writing $X=\sum_{m\in\Mset}b_{m}X_{m}$, we find that
\begin{equation}
E\left(X-E\left[X|\left\{ Y_{m}\right\} _{m\in\Mset}\right]\right)^{2}=\sum_{m\in\Mset}b_{m}^{2}E\left(X_{m}-E\left[X_{m}|Y_{m}\right]\right)^{2}=\sum_{m\in\Mset}b_{m}^{2}\textrm{mmse}_{X_{m}}(\frac{\gamma}{\sigma_{m}^{2}})
\end{equation}

\end{IEEEproof}
Specializing to $\sigma_{m}^{2}=1/\sum_{k\in\Kset}a_{k}^{2}$ for
any $m\in\Mset$ yields the bound,
\begin{equation}
\textrm{mmse}_{X}(\gamma)\geq\sum_{m=1}^{M}b_{m}^{2}\textrm{mmse}_{X_{m}}\left(\gamma\right)\label{eq:mmse-genie-same-sigma}
\end{equation}
Specializing further to $\Pset_{k}=\left\{ k\right\} $ for any $k\in\Kset$
($\Mset=\Kset$) yields the very simple bound
\begin{equation}
\textrm{mmse}_{X}(\gamma)\geq\textrm{mmse}_{x}\left(\gamma\right)\label{eq:mmse-genie-singleton-clusters}
\end{equation}

Another interesting choice is $\sigma_{m'}^{2}=1/b_{m'}^{2}$ for
some $m'$ and $\sigma_{m}^{2}=0$ for all $m\neq m'$. In this case,
\begin{equation}
\textrm{mmse}_{X}(\gamma)\geq b_{m'}^{2}\textrm{mmse}_{X_{m}}\left(b_{m'}^{2}\gamma\right)\label{eq:mmse-genie-one-cluster}
\end{equation}
Applying (\ref{eq:mmse-genie-same-sigma}) to $\textrm{mmse}_{X_{m}}\left(\cdot\right)$
in (\ref{eq:mmse-genie-one-cluster}) yields the simpler bound,
\begin{equation}
\textrm{mmse}_{X}(\gamma)\geq b_{m'}^{2}\textrm{mmse}_{x}\left(b_{m'}^{2}\gamma\right)\label{eq:mmse-genie-one-cluster-simplified}
\end{equation}

Note that the smaller $b_{m'}$ is, the tighter the bound in high
SNR's, while the opposite is true for low SNR's.

\subsection{Information-Estimation application}

In this subsection we will rely on the definitions of $\mu_{0},\mu_{1}$
and $\beta_{0},\beta_{1}$ as given in (\ref{eq:mu01_def}) and (\ref{eq:beta01_def}),
respectively. We also define
\begin{equation}
S\triangleq\frac{P_{x}}{E{\hat{m}}^{2}}\label{eq:S_def}
\end{equation}
with $P_{x}$ the input power and $\hat{m}$ the Gaussian noise component
resulting from the application of the unbiased MMSE-DFE on the channel
output. Explicit expressions for $S,\beta_{1}$ and $\beta_{0}$ in
terms of the ISI channel transfer function $H\left(\theta\right)$
are given in Appendix \ref{app:second-order-expressions}, in (\ref{eq:S expression}),
(\ref{eq:beta1 expression}) and (\ref{eq:beta0 expression}), respectively.
Under this notation, are bound is given in the following,
\begin{thm}
For any input distribution and any ISI channel,
\begin{equation}
\Immse\geq I_{\mathrm{IE}}\left(\gamma_{1},\gamma_{2}\right)\triangleq I_{x}(\beta_{0}^{2}\gamma_{1})-I_{x}\left(\gamma_{1}\right)+I_{x}\left(\gamma_{2}\right)-\frac{1}{2}\log\left(1+\beta_{1}^{2}\gamma_{2}\right)\label{eq:ie_bound}
\end{equation}
For any $0\leq\gamma_{1}\leq\gamma_{2}\leq S$.\end{thm}
\begin{IEEEproof}
As is the proof of Theorem \ref{thm:slc_wrong} we rewrite $\Immse$
as
\begin{equation}
\Immse=I\left(\mu_{0}\:;\:\mu_{0}+\hat{m}\right)-I\left(\mu_{1}\:;\:\mu_{1}+\hat{m}\right)\label{eq:Immse_as_2_part_diff}
\end{equation}
We use the Guo-Shamai-Vedú theorem \cite{guo2005mutual} to write
\begin{equation}
2I\left(\mu_{0}\:;\:\mu_{0}+\hat{m}\right)=\int_{0}^{\beta_{0}^{2}\gamma_{1}}+\int_{\beta_{0}^{2}\gamma_{1}}^{\beta_{0}^{2}\gamma_{2}}+\int_{\beta_{0}^{2}\gamma_{2}}^{\beta_{0}^{2}S}\textrm{mmse}_{\bar{\mu}_{0}}\left(\gamma\right)d\gamma\label{eq:Imu0_as_3_path_int}
\end{equation}
\begin{equation}
2I\left(\mu_{1}\:;\:\mu_{1}+\hat{m}\right)=\int_{0}^{\beta_{1}^{2}\gamma_{2}}+\int_{\beta_{1}^{2}\gamma_{2}}^{\beta_{1}^{2}S}\textrm{mmse}_{\bar{\mu}_{1}}\left(\gamma\right)d\gamma\label{eq:Imu1_as_2_part_int}
\end{equation}
where $\bar{\mu}_{i}=\mu_{i}/\beta_{i}=\sum_{k\geq i}\alpha_{k}x_{k}/\beta_{i}$
is scaled to unit power. Since $\bar{\mu}_{0}$ is a unit power sum
of scaled i.i.d. random variables, Lemma \ref{lem:mmse_iid_sum_bound}
and the bounds derived from it apply. For $0\leq\gamma\leq\beta_{0}^{2}\gamma_{1}$,
we apply the bound (\ref{eq:mmse-genie-singleton-clusters}) to obtain,
\begin{equation}
\int_{0}^{\beta_{0}^{2}\gamma_{1}}\textrm{mmse}_{\bar{\mu}_{0}}\left(\gamma\right)d\gamma\geq\int_{0}^{\beta_{0}^{2}\gamma_{1}}\textrm{mmse}_{\bar{x}}\left(\gamma\right)d\gamma=2I_{x}\left(\beta_{0}^{2}\gamma_{1}\right)\label{eq:mmse_mu0_b1}
\end{equation}
where $\bar{x}=x/\sqrt{P_{x}}$.

For $\beta_{0}^{2}\gamma_{1}\leq\gamma\leq\beta_{0}^{2}\gamma_{2}$,
we apply (\ref{eq:mmse-genie-one-cluster-simplified}) with only $x_{0}$
is the chosen subset, i.e. $\Pset_{m'}=\left\{ 0\right\} $, yielding
\begin{equation}
\int_{\beta_{0}^{2}\gamma_{1}}^{\beta_{0}^{2}\gamma_{2}}\textrm{mmse}_{\bar{\mu}_{0}}\left(\gamma\right)d\gamma\geq\int_{\beta_{0}^{2}\gamma_{1}}^{\beta_{0}^{2}\gamma_{2}}\frac{1}{\beta_{0}^{2}}\textrm{mmse}_{\bar{x}}\left(\frac{1}{\beta_{0}^{2}}\gamma\right)d\gamma=2I_{x}\left(\gamma_{2}\right)-2I_{x}\left(\gamma_{1}\right)\label{eq:mmse_mu0_b2}
\end{equation}
since $b_{m'}^{2}=\alpha_{0}^{2}/\beta_{0}^{2}=1/\beta_{0}^{2}$.

For $\beta_{0}^{2}\gamma_{2}\leq\gamma\leq\beta_{0}^{2}S$, we apply
(\ref{eq:mmse-genie-one-cluster}) with the chosen subsets including
all indices but $0$, i.e. $\Pset=\Kset\setminus\left\{ 0\right\} $,
yielding
\begin{equation}
\int_{\beta_{0}^{2}\gamma_{2}}^{\beta_{0}^{2}S}\textrm{mmse}_{\bar{\mu}_{0}}\left(\gamma\right)d\gamma\geq\int_{\beta_{0}^{2}S\gamma_{2}}^{\beta_{0}^{2}S}\frac{\beta_{1}^{2}}{\beta_{0}^{2}}\textrm{mmse}_{\bar{\mu}_{1}}\left(\frac{\beta_{1}^{2}}{\beta_{0}^{2}}\gamma\right)d\gamma=\int_{\beta_{1}^{2}\gamma_{2}}^{\beta_{1}^{2}S}\textrm{mmse}_{\bar{\mu}_{1}}\left(\gamma\right)d\gamma\label{eq:mmse_mu0_b3}
\end{equation}

Finally, applying the Gaussian upper bound $\textrm{mmse}_{\bar{\mu}_{1}}\left(\gamma\right)\leq1/\left(1+\gamma\right)$
we have
\begin{equation}
\int_{0}^{\beta_{1}^{2}\gamma_{2}}\textrm{mmse}_{\bar{\mu}_{1}}\left(\gamma\right)d\gamma\leq\log\left(1+\beta_{1}^{2}\gamma_{2}\right)\label{eq:mmse_mu1_b}
\end{equation}
Substituting (\ref{eq:mmse_mu0_b1}), (\ref{eq:mmse_mu0_b2}) and
(\ref{eq:mmse_mu0_b3}) to (\ref{eq:Imu0_as_3_path_int}), (\ref{eq:mmse_mu1_b})
to (\ref{eq:Imu1_as_2_part_int}), and (\ref{eq:Imu0_as_3_path_int})
and (\ref{eq:Imu1_as_2_part_int}) into (\ref{eq:Immse_as_2_part_diff})
yields the desired result.
\end{IEEEproof}
Choosing $\gamma_{1}=\gamma_{2}=S$ yields the following simple bound
\begin{equation}
\Immse\geq I_{\mathrm{IE},simple}\triangleq I_{x}\left(\beta_{0}^{2}S\right)-\frac{1}{2}\log\left(1+\beta_{1}^{2}S\right)
\end{equation}
Note that $\frac{1}{2}\log\left(1+\gamma\right)=I_{g}\left(\gamma\right)$
is the Gaussian mutual information function. Also note that the bound
holds with equality for Gaussian inputs, i.e. $\Iach_{g}=I_{g}\left(\beta_{0}^{2}S\right)-I_{g}\left(\beta_{1}^{2}S\right)$,
where $\Iach_{g}$ is the i.i.d. channel capacity given in (\ref{eq:iid capacity}).
The bound $I_{\mathrm{IE},simple}$ simple is guaranteed to be tight
for sufficiently low SNR's, since $I_{x}\left(\gamma\right)\approx I_{g}\left(\gamma\right)\approx\gamma/2$
as $\gamma$ tends to zero. It also guaranteed to be tight for sufficiently
high SNR's, since $\beta_{0}^{2}S\to\infty$ and $\beta_{1}^{2}S\to0$
and $P_{x}/N_{0}\to\infty$, ensuring that $I_{\mathrm{IE},simple}$
converges to the input entropy.

The above discussion leads us to conjecture that
\begin{equation}
I_{\mathrm{IE},conj}\triangleq I_{x}\left(\beta_{0}^{2}S\right)-I_{x}\left(\beta_{1}^{2}S\right)
\end{equation}
is also a lower bound for $\Immse$. This is equivalent to conjecturing
that the bound $I\left(\mu_{0}\:;\:\mu_{0}+\hat{m}\right)\geq I_{x}\left(\beta_{0}^{2}S\right)$
is always looser than the bound $I\left(\mu_{1}\:;\:\mu_{1}+\hat{m}\right)\geq I_{x}\left(\beta_{1}^{2}S\right)$
, i.e. that
\[
I\left(\mu_{0}\:;\:\mu_{0}+\hat{m}\right)-I_{x}\left(\beta_{0}^{2}S\right)\geq I\left(\mu_{1}\:;\:\mu_{1}+\hat{m}\right)-I_{x}\left(\beta_{1}^{2}S\right)
\]
Attempts to prove this conjectured bound were so far unsuccessful.
In all simulations preformed, $I_{\mathrm{IE},conj}$ never exceeded
$\Immse$, supporting this conjecture. However, for common input distributions
such as BPSK, $I_{\mathrm{IE},conj}$ was seen to offer very little
improvement over $I_{\mathrm{IE},simple}$, even for channels with
severe ISI. This is due to the fact that generally, $E\mu_{1}^{2}$
is of the order of $E{\hat{m}}^{2}$ or smaller, and therefore the
mutual information $I\left(\mu_{1}\:;\:\mu_{1}+\hat{m}\right)$ is
very well approximated by the Gaussian upper bound. In the simulations
described in the following subsection, the difference between $I_{\mathrm{IE},conj}$
and $I_{\mathrm{IE},simple}$ was not noticeable and therefore only
$I_{\mathrm{IE},simple}$ was plotted.

Let
\begin{equation}
\Immse\geq I_{\mathrm{IE},opt}\triangleq\max_{0\leq\gamma_{1}\leq\gamma_{2}\leq S}I_{\mathrm{IE}}\left(\gamma_{1},\gamma_{2}\right)
\end{equation}
 be the optimal lower bound, and let $\gamma_{1}^{*}$ and $\gamma_{2}^{*}$
be its optimizers. The values $\gamma_{1}^{*}$, $\gamma_{2}^{*}$
can be easily determined using the following procedure. If $\beta_{0}^{2}\textrm{mmse}_{\bar{x}}\left(\beta_{0}^{2}S\right)\geq\textrm{mmse}_{\bar{x}}\left(S\right)$,
then $\gamma_{1}^{*}=\gamma_{2}^{*}=S$ and $I_{\mathrm{IE},opt}=I_{\mathrm{IE},simple}$.
Otherwise, $\gamma_{1}^{*}$ satisfies
\begin{equation}
\beta_{0}^{2}\textrm{mmse}_{\bar{x}}\left(\beta_{0}^{2}\gamma_{1}^{*}\right)=\textrm{mmse}_{\bar{x}}\left(\gamma_{1}^{*}\right)
\end{equation}
Next, if $\textrm{mmse}_{\bar{x}}\left(S\right)\geq\beta_{1}^{2}/\left(1+\beta_{1}^{2}S\right)$,
then $\gamma_{2}^{*}=S$. Otherwise, $\gamma_{2}^{*}$ satisfies
\begin{equation}
\textrm{mmse}_{\bar{x}}\left(\gamma_{2}^{*}\right)=\frac{\beta_{1}^{2}}{1+\beta_{1}^{2}\gamma_{2}^{*}}
\end{equation}

We conclude by noting that the bound (\ref{eq:ie_bound}) can be further
generalized using Lemma \ref{lem:mmse_iid_sum_bound} by using other
partitions of the ISI taps and adding further degrees of freedom to
the optimization. However, numerical experimentation indicates that
such generalizations offer very little improvement in tightness.

\subsection{Evaluation of the bounds}

\begin{figure}
\centering{}\includegraphics[width=14cm]{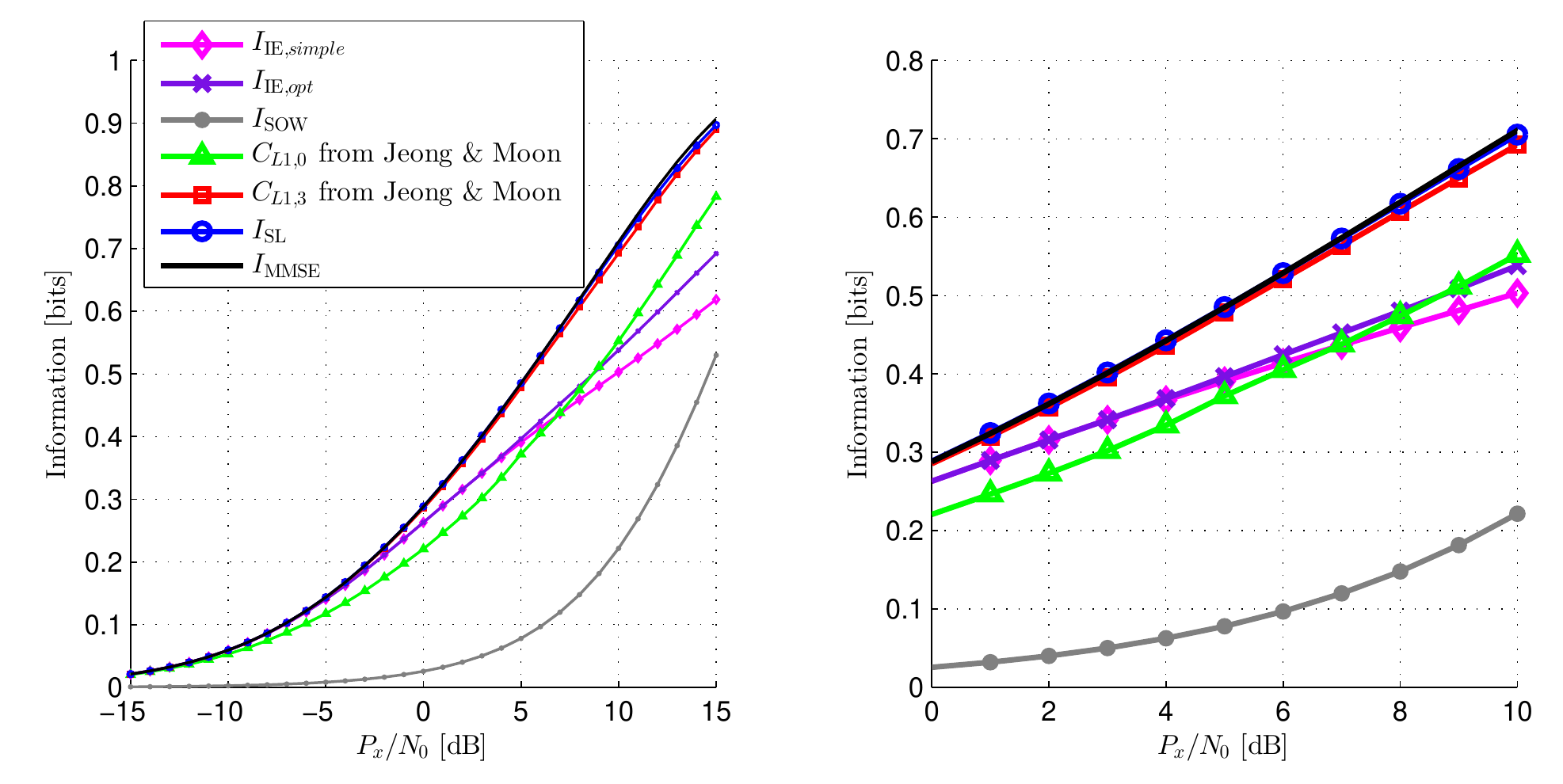}\caption{\label{fig:Iapprox_compare}Comparison of bounds for $\Immse$ for
BPSK input and ISI channel \protect{(\ref{eq:Jeong_h})}}
\end{figure}

Figure \ref{fig:Iapprox_compare} plots $\Immse$, $\Isl$, the Shamai-Ozarow-Wyner
bound $\Isow$ (\ref{eq:SOW bound}) and the bounds $I_{\mathrm{IE},opt}$
and $I_{\mathrm{IE},simple}$, for BPSK input the most severe example
ISI channel that appeared in \cite{jeong2012easily},
\begin{equation}
h=\left[0.19,\,0.35,\,0.46,\,0.5,\,0.46,\,0.35,\,0.19\right]\label{eq:Jeong_h}
\end{equation}
The lower bounds $C_{L1,0}$ and $C_{L1,3}$ proposed in \cite{jeong2012easily}
are also plotted. $\Immse$ has been approximated using the procedure
described in Section \ref{sec:counterexamples}.

It is seen that the proposed bounds $I_{\mathrm{IE},opt}$ and $I_{\mathrm{IE},simple}$
are identical and tight in the low SNR region. For medium to high
SNR's, $I_{\mathrm{IE},opt}$ improves on $I_{\mathrm{IE},simple}$,
but both bounds are not very tight. Additionally, $\Isow$ is very
loose in this example, reflecting the poor performance of decision
feedback zero-forcing equalization in severe ISI conditions. In comparison
to the bounds from \cite{jeong2012easily}, $I_{\mathrm{IE},opt}$
and $I_{\mathrm{IE},simple}$ are tighter than the simple single-letter
bound $C_{L1,0}$ in the low to medium SNR region, but are less tight
for higher SNR's. The tightened 3-letter bound $C_{L1,3}$ is tighter
than our proposed bound for all SNR's.

\begin{figure}
\centering{}\includegraphics[width=14.5cm]{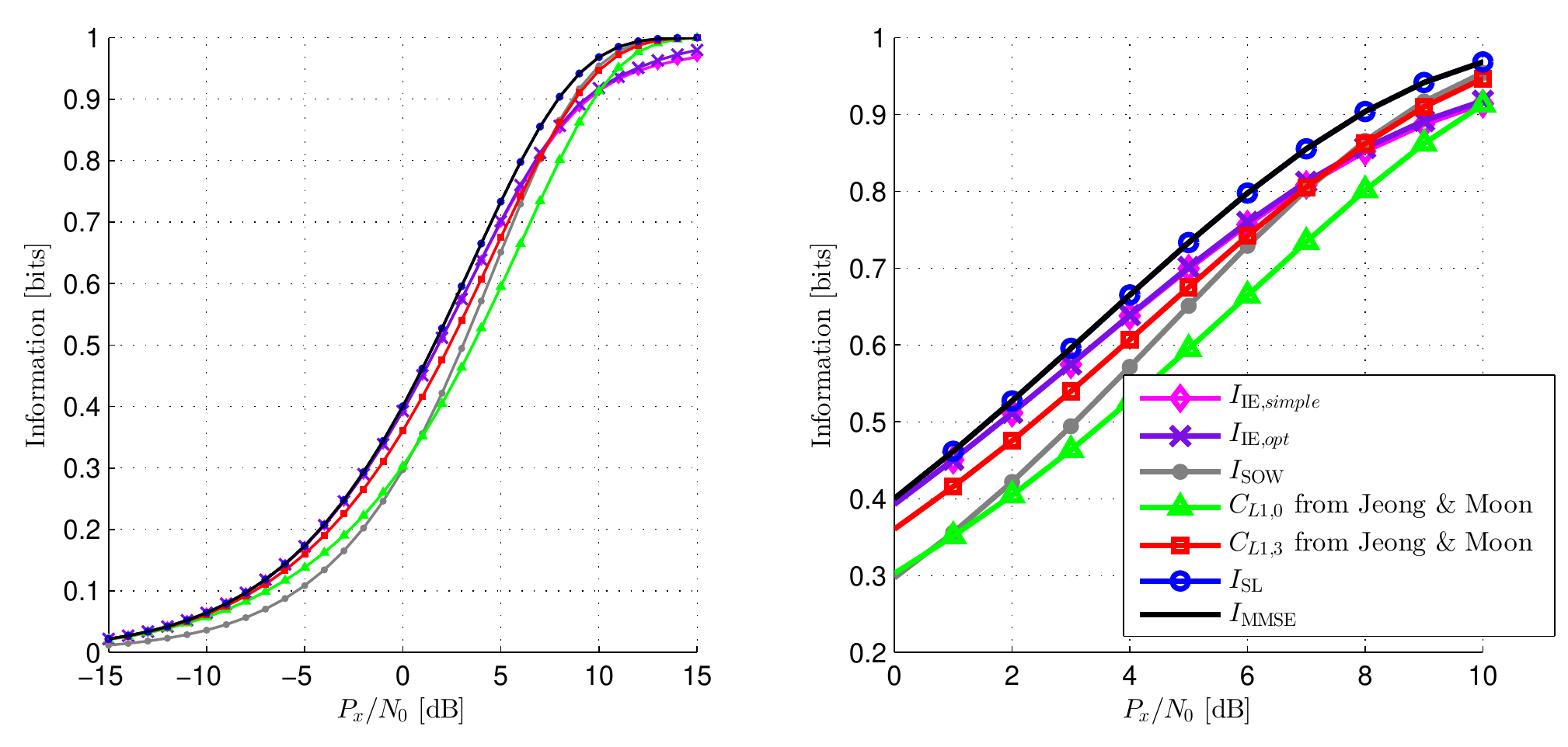}\caption{\label{fig:Iapprox_compare-2}Comparison of bounds for $\Immse$ for
BPSK input and ISI channel \protect{(\ref{eq:Jeong_h_spaced})}}
\end{figure}

As reported in \cite{jeong2012easily}, the bounds reported there
become looser when there is no small set of dominant coefficients
in the residual ISI sequence $\alpha_{1}^{\infty}$, as often happens
in highly scattered multipath channels. In order to simulate such
channel, the impulse response (\ref{eq:Jeong_h}) was spaced by adding
3 and 5 null taps around the main tap, yielding,
\begin{equation}
\tilde{h}=[0.19,\,0.35,\,0.46,\,\underset{3}{\underbrace{0\cdots0}},\,0.5,\,\underset{5}{\underbrace{0\cdots0}},\,0.46,\,0.35,\,0.19]\label{eq:Jeong_h_spaced}
\end{equation}
The experiment described above was repeated with the modified ISI
channel (\ref{eq:Jeong_h_spaced}), and the results are shown in Figure
\ref{fig:Iapprox_compare-2}. It is seen that here the bounds of \cite{jeong2012easily}
are considerably looser, while our proposed bounds retain their tightness.
The bound in \cite{jeong2012easily} may be tightened by increasing
the parameter $M$ beyond 3, but at the cost of an exponentially increasing
computational load and loss of analytic tractability. It is interesting
to note that the spacing of the ISI channel has actually reduced the
severity of the ISI --- this is evident from the higher information
rates attained, as well as from the improvement in tightness of the
Shamai-Ozarow-Wyner bound. In fact, in this experiment $\Isow$ proved
to be the tightest bound in high SNR's.

As a last point of comparison we remark that our bounds apply to any
input distribution, while the bounds of \cite{jeong2012easily} are
developed only for symmetric binary input.

\section{\label{sec:conc}Conclusion}

This paper addressed the long-standing Shamai-Laroia conjecture from
several directions. First, the original conjecture was shown analytically
not to hold. Next, a natrual relaxation of the conjecture was considered,
in which $\Isl$ is proposed as a lower bound for $\Iach$, the single-carrier
achievable rate. It was shown by means of Monte-Carlo simulation that
this weakened conjecture does not hold as well, by means of a counterexample
based on a highly skewed binary input. A positive result on the relaxed
conjecture $\Iach\geq\Isl$ is then presented, showing that it holds
in the high SNR regime. Finally, alternative bounds for the achievable
rate are proven. While not as tight as $\Isl$, these bounds have
expressions nearly as simple.

Enabling all of our results are recently discovered properties of
the mutual information in the scalar additive Gaussian channel with
arbitrarily distributed inputs. Namely, the low SNR power series of
\cite{guo2011estimation} and the Guo-Shamai-Verdú Information-Estimation
relation \cite{guo2005mutual} find useful application in this work.

Both the negative and positive results in this paper are of practical
relevance, as $\Isl$ is an often used approximation for the achievable
rate in the ISI channel. On the one hand, we disprove the conjecture
that $\Isl$ is a lower bound to the achievable rate, invoking caution
when it is used as such. On the other hand, our high-SNR proof that
$\Iach\geq\Isl$ helps to theoretically establish $\Isl$ as a good
approximation.

A remaining open question is whether the inequality $\Iach\geq\Isl$
is true for all SNR's for commonly used input distributions such as
PSK or QAM. While numeric experimentation supports this refined conjecture,
no theoretical proof is known. This question is of particular interest
in the context of comparison between the achievable rates of OFDM
and single-carrier modulation in the ISI channel, where a fixed i.i.d.
input distribution is assumed. In this setting, $\Isl$ can be shown
to essentially act as an \emph{upper bound} for the OFDM achievable
rate \cite{carmon2013comparison}. Thus, proving that $\Iach\geq\Isl$
for a given input distribution is tantamount to showing that the single-carrier
achievable rate is superior to that of OFDM, regardless of the specific
ISI channel, as long as that distribution is used.

\section*{Acknowledgment}

The authors wish to thank Tsachy Weissman for helpful discussions.

\bibliographystyle{unsrt}

\appendices{}

\section{\label{app:second-order-expressions}Noise and interference variance
in the MMSE-DFE}

In this section we derive closed form expressions for the quantities
$\beta_{0},\beta_{1},\epsilon_{0}$ and $S$ defined in equations
(\ref{eq:beta01_def}), (\ref{eq:epsilon01_def}) and (\ref{eq:S_def}),
respectively. The expressions are given in terms of the output SNRs
of the linear and decision-feedback MMSE equalizers, which in turn
admit simple expressions in terms of the ISI channel transfer function
$H\left(\theta\right)=\sum_{n}h_{n}e^{-jn\theta}$.

As in (\ref{eq:Immse-def}), the output of the unbiased MMSE at samples
0 given by
\[
z_{0}=x_{0}+\sum_{k\geq1}\hat{\alpha}_{k}x_{k}+\hat{m}
\]
where $x_{-\infty}^{\infty}$ is the channel input sequence, ${\hat{\alpha}}_{1}^{\infty}$
are the residual ISI coefficients and $\hat{m}$ is an independent
Gaussian noise component. The values of ${\hat{\alpha}}_{1}^{\infty}$
and $E{\hat{m}}^{2}$ maximize
\begin{equation}
\frac{P_{x}}{E\left(\sum_{k\geq1}\hat{\alpha}_{k}x_{k}+\hat{m}\right)^{2}}=\SNR[DFE-U]=\SNR[DFE]-1\label{eq:SNR-dfe expression appendix}
\end{equation}
where
\begin{gather}
\SNR[LE]=\left[\frac{1}{2\pi}{\displaystyle \intop_{-\pi}^{\pi}\frac{d\theta}{1+\frac{P_{x}}{N_{0}}\left|H(\theta)\right|^{2}}}\right]^{-1}\label{eq:LE SNR def}\\
\SNR[DFE]=\exp\left\{ \frac{1}{2\pi}{\displaystyle \intop_{-\pi}^{\pi}\log\left[1+\frac{P_{x}}{N_{0}}\left|H(\theta)\right|^{2}\right]d\theta}\right\} \label{eq:DFE SNR def}
\end{gather}
stand for SNR of the (biased) linear and decision-feedback equalizers,
respectively

An explicit expression for the output of the biased MMSE-DFE is given
in equation (50) of \cite{cioffi1995mmse}, from which it can be read
that the PSD of the Gaussian noise component is given by
\[
\left(\frac{P_{x}/N_{0}}{\SNR[DFE]}\right)^{2}\frac{\SNR[DFE]N_{0}\left|H\left(\theta\right)\right|^{2}}{1+\frac{P_{x}}{N_{0}}\left|H\left(\theta\right)\right|^{2}}=\left(\frac{P_{x}}{\SNR[DFE]}\right)\frac{\frac{P_{x}}{N_{0}}\left|H\left(\theta\right)\right|^{2}}{1+\frac{P_{x}}{N_{0}}\left|H\left(\theta\right)\right|^{2}}
\]
It is also shown in \cite{cioffi1995mmse} that the unbiased MMSE-DFE
is obtained by scaling the output of the MMSE-DFE by a factor of $\SNR[DFE]/\left(\SNR[DFE]-1\right)$.
Combining these expressions it seen that,
\begin{eqnarray}
E{\hat{m}}^{2} & = & P_{x}\frac{\SNR[DFE]}{\left(\SNR[DFE]-1\right)^{2}}\frac{1}{2\pi}{\displaystyle \intop_{-\pi}^{\pi}\left[1-\frac{1}{1+\frac{P_{x}}{N_{0}}\left|H\left(\theta\right)\right|^{2}}\right]}d\theta\nonumber \\
 & =P_{x} & \frac{\SNR[DFE]}{\left(\SNR[DFE]-1\right)^{2}}\frac{\SNR[LE]-1}{\SNR[LE]}\label{eq:Em2 expression}
\end{eqnarray}
and
\begin{equation}
S=\frac{P_{x}}{E{\hat{m}}^{2}}=\frac{\left(\SNR[DFE]-1\right)^{2}}{\SNR[DFE]}\frac{\SNR[LE]}{\SNR[LE]-1}\label{eq:S expression}
\end{equation}

Noticing that $E\left(\sum_{k\geq1}\hat{\alpha}_{k}x_{k}+\hat{m}\right)^{2}=\beta_{1}^{2}P_{x}+E{\hat{m}}^{2}$
and plugging (\ref{eq:Em2 expression}) into (\ref{eq:SNR-dfe expression appendix}),
we find that
\begin{equation}
\beta_{1}^{2}=\frac{1}{\SNR[DFE]-1}-\frac{E{\hat{m}}^{2}}{P_{x}}=\frac{\SNR[DFE]/\SNR[LE]-1}{\left(\SNR[DFE]-1\right)^{2}}\label{eq:beta1 expression}
\end{equation}
and hence
\begin{equation}
\beta_{0}^{2}=1+\beta_{1}^{2}=\frac{\SNR[DFE]}{\SNR[DFE]-1}\left[1-\left(\SNR[LE]\right)^{-1}\frac{\SNR[LE]-1}{\SNR[DFE]-1}\right]\label{eq:beta0 expression}
\end{equation}

Combining the above results we find that
\begin{equation}
\epsilon_{1}=\frac{P_{x}\beta_{1}^{2}}{E{\hat{m}}^{2}}=\frac{1-\SNR[LE]/\SNR[DFE]}{\SNR[LE]-1}\label{eq:eps1 expression}
\end{equation}
and
\begin{equation}
\epsilon_{0}=\frac{P_{x}\beta_{0}^{2}}{E{\hat{m}}^{2}}=\SNR[LE]\frac{\SNR[DFE]-1}{\SNR[LE]-1}-1\label{eq:eps0 expression}
\end{equation}

\end{document}